\batchmode
\makeatletter
\def\input@path{{\string"/Users/hari/Git/private/phd/papers/Working/A Numerical Scheme for Non-Newtonian Fluids and Plastic Solids under the GPR Model/\string"}}
\makeatother
\documentclass[twoside,english,3p]{elsarticle}
\usepackage[T1]{fontenc}
\usepackage[latin9]{inputenc}
\usepackage{prettyref}
\usepackage{mathtools}
\usepackage{amsmath}
\usepackage{graphicx}
\usepackage{wasysym}
\usepackage{nomencl}

\providecommand{\makenomenclature}{\makeglossary}
\makenomenclature

\makeatletter
\journal{Journal of Computational Physics}

\usepackage{fancyhdr}
\makeatother

\usepackage{babel}
\begin{document}

\begin{frontmatter}{}

\title{A Numerical Scheme for Non-Newtonian Fluids and Plastic Solids under
the GPR Model}

\author[rvt]{Haran Jackson\corref{cor1}\corref{cor2}}

\ead{hj305@cam.ac.uk}

\author[rvt]{Nikos Nikiforakis}

\cortext[cor1]{Corresponding author}

\cortext[cor2]{Principal corresponding author}

\address[rvt]{Cavendish Laboratory, Department of Physics, Cambridge University,
UK}
\begin{abstract}
A method for modeling non-Newtonian fluids (dilatants and pseudoplastics)
by a power law under the Godunov-Peshkov-Romenski model is presented,
along with a new numerical scheme for solving this system. The scheme
is also modified to solve the corresponding system for power-law elastoplastic
solids.

The scheme is based on a temporal operator splitting, with the homogeneous
system solved using a finite volume method based on a WENO reconstruction,
and the temporal ODEs solved using an analytical approximate solution.
The method is found to perform favorably against problems with known
exact solutions, and numerical solutions published in the open literature.
It is simple to implement, and to the best of the authors' knowledge
it is currently the only method for solving this modified version
of the GPR model.
\end{abstract}

\end{frontmatter}{}

\global\long\def\dev{\operatorname{dev}}

\global\long\def\tr{\operatorname{tr}}

\global\long\def\diag{\operatorname{diag}}

\section{Introduction}

\subsection{Background}

The Godunov-Pehskov-Romenski (GPR) model of continuum mechanics (see
\citet{peshkov_hyperbolic_2016}) has been purported to represent
an alternative formulation to describe both fluids and solids within
the same hyperbolic system of differential equations. From a practical
perspective, the potential ramifications of this include: the simplification
of software made for the simulation of phenomena involving different
states of matter (as commented on in \citet{jackson_fast_2017});
and the use of the vast array of effective numerical solvers designed
for first-order hyperbolic systems. From a theoretical perspective,
an advantage of the GPR model is that it cannot produce waves of infinite
speed, unlike the parabolic Navier-Stokes equations. Additionally,
the first-principles derivation of the mechanism by which viscous
effects appear under the GPR model has been commented to be more appropriate
than the more phenomenological viscous law appearing in the traditional
Navier-Stokes formulation (see \citet{peshkov_unified_2017}).

Thus far, the GPR model has been solved for a wide array of different
fluids (inviscid and viscous Newtonian) and solids (elastic and elastoplastic)
(see \citet{dumbser_high_2015,boscheri_cell_2016,peshkov_theoretical_2018,peshkov_unified_2017}).
It has also been extended to incorporate the effects of electrodynamics
(see \citet{dumbser_high_2016}) and general relativity (see \citet{peshkov_general_2018}).
To the best of our knowledge, it is yet to be formally extended to
include non-Newtonian power-law fluids, however. A method for doing
so is presented in this paper. Building upon the work of \citet{jackson_fast_2017}
(in which a numerical scheme based upon a split solver was presented
for Newtonian fluids and elastic solids under the GPR model) a numerical
scheme is then presented in \prettyref{sec:Numerical-Schemes} for
solving this new model. This scheme is adapted to work also for elastoplastic
power-law materials. The scheme is validated against several 1D and
2D tests in \prettyref{sec:Numerical-Results}, with discussion presented
in \prettyref{sec:Conclusions}.

\subsection{The GPR Model\label{subsec:The-GPR-Model}}

The GPR model, first introduced in \citet{peshkov_hyperbolic_2016}
- and expanded upon by \citet{dumbser_high_2015} and \citet{boscheri_cell_2016}
- takes the following form:

\begin{subequations}

\begin{align}
\frac{\partial\rho}{\partial t}+\frac{\partial\left(\rho v_{k}\right)}{\partial x_{k}} & =0\label{eq:DensityEquation}\\
\frac{\partial\left(\rho v_{i}\right)}{\partial t}+\frac{\partial(\rho v_{i}v_{k}+p\delta_{ik}-\sigma_{ik})}{\partial x_{k}} & =0\label{eq:MomentumEquation}\\
\frac{\partial A_{ij}}{\partial t}+\frac{\partial\left(A_{ik}v_{k}\right)}{\partial x_{j}}+v_{k}\left(\frac{\partial A_{ij}}{\partial x_{k}}-\frac{\partial A_{ik}}{\partial x_{j}}\right) & =-\frac{\psi_{ij}}{\theta_{1}}\label{eq:DistortionEquation}\\
\frac{\partial\left(\rho J_{i}\right)}{\partial t}+\frac{\partial\left(\rho J_{i}v_{k}+T\delta_{ik}\right)}{\partial x_{k}} & =-\frac{\rho H_{i}}{\theta_{2}}\label{eq:ThermalEquation}\\
\frac{\partial\left(\rho E\right)}{\partial t}+\frac{\partial\left(\rho Ev_{k}+\left(p\delta_{ik}-\sigma_{ik}\right)v_{i}+q_{k}\right)}{\partial x_{k}} & =0\label{eq:EnergyEquation}
\end{align}

\end{subequations}

\nomenclature[g]{$\rho$}{Density}\nomenclature[a]{$\boldsymbol{v}$}{Velocity}\nomenclature[a]{$p$}{Pressure}\nomenclature[g]{$\delta$}{Kronecker delta}\nomenclature[g]{$\sigma$}{Viscous shear stress tensor}\nomenclature[a]{$A$}{Distortion tensor}\nomenclature[a]{$\boldsymbol{J}$}{Thermal impulse vector}\nomenclature[a]{$T$}{Temperature}\nomenclature[a]{$s$}{Entropy}\nomenclature[g]{${\tau}_{1}$}{Strain dissipation time}\nomenclature[g]{${\tau}_{2}$}{Thermal impulse dissipation time}\nomenclature[a]{$E$}{Total specific energy}\nomenclature[a]{$\boldsymbol{q}$}{Heat flux vector}\nomenclature[a]{$x$}{Space variable}\nomenclature[a]{$t$}{Time variable}

where $\theta_{1}$ and $\theta_{2}$ are positive scalar functions,
and $\psi=\frac{\partial E}{\partial A}$ and $\boldsymbol{H}=\frac{\partial E}{\partial\boldsymbol{J}}$.
The following definitions are given:

\begin{subequations}

\begin{align}
p & =\rho^{2}\left.\frac{\partial E}{\partial\rho}\right|_{s,A}\label{eq:p def}\\
\sigma & =-\rho A^{T}\left.\frac{\partial E}{\partial A}\right|_{\rho,s}\label{eq:sig def}\\
T & =\left.\frac{\partial E}{\partial s}\right|_{\rho,A}\label{eq:T def}\\
\boldsymbol{q} & =T\frac{\partial E}{\partial\boldsymbol{J}}\label{eq:q def}
\end{align}

\end{subequations}

To close the system, the EOS must be specified, from which the above
quantities and the sources can be derived. $E$ is the sum of the
contributions of the energies at the molecular scale (microscale),
the material element\footnote{The concept of a \textit{material element} corresponds to that of
a fluid parcel from fluid dynamics, applied to both fluids and solids.} scale (mesoscale), and the flow scale (macroscale):

\nomenclature[z]{EOS}{Equation of State}

\begin{equation}
E=E_{1}\left(\rho,s\right)+E_{2}\left(\rho,s,A,\boldsymbol{J}\right)+E_{3}\left(\boldsymbol{v}\right)\label{eq:EnergyDefinition}
\end{equation}

Here, as in previous studies, such as \citet{dumbser_high_2015} and
\citet{boscheri_cell_2016}, $E_{1}$ is taken to be either the ideal
gas EOS, a shock Mie-Gruneisen EOS, or the EOS of nonlinear hyperelasticity
(see \citet{barton_exact_2009}).

$E_{2}$ has the following quadratic form:

\begin{equation}
E_{2}=\frac{c_{s}\left(\rho,s\right)^{2}}{4}\left\Vert \dev\left(G\right)\right\Vert _{F}^{2}+\frac{c_{t}\left(\rho,s\right)^{2}}{2}\left\Vert \boldsymbol{J}\right\Vert ^{2}
\end{equation}

\nomenclature[a]{$c_s$}{Characteristic velocity of transverse perturbations}\nomenclature[x]{$\left\Vert \cdot\right\Vert $}{Euclidean vector norm}\nomenclature[x]{${\left\Vert \cdot\right\Vert}_F $}{Frobenius matrix norm}\nomenclature[g]{$c_{t}$}{Coefficient in the thermal impulse contribution to the energy (elsewhere denoted by $\alpha$)}

$c_{s}$ is the characteristic velocity of transverse perturbations.
$c_{t}$ is related to the characteristic velocity of propagation
of heat waves\footnote{Note that \citet{dumbser_high_2015} denotes this variable by $\alpha$,
which is avoided here due to a clash with a parameter of one of the
equations of state used.}:

\begin{equation}
c_{h}=\frac{c_{t}}{\rho}\sqrt{\frac{T}{c_{v}}}
\end{equation}

\nomenclature[a]{$c_h$}{Characteristic velocity of heat waves}

In previous studies, $c_{t}$ has been taken to be constant, as it
will be in this study.

$G=A^{T}A$ is the Gramian matrix of the distortion tensor, and $\dev\left(G\right)$
is the deviator (trace-free part) of $G$:

\begin{equation}
\dev\left(G\right)=G-\frac{1}{3}\tr\left(G\right)I
\end{equation}

$E_{3}$ is the usual specific kinetic energy per unit mass:

\begin{equation}
E_{3}=\frac{1}{2}\left\Vert \boldsymbol{v}\right\Vert ^{2}
\end{equation}

The following forms are taken:

\begin{subequations}

\begin{align}
\theta_{1} & =\frac{\tau_{1}c_{s}^{2}}{3\left|A\right|^{\frac{5}{3}}}\\
\theta_{2} & =\tau_{2}c_{t}^{2}\frac{\rho T_{0}}{\rho_{0}T}
\end{align}

\end{subequations}

\begin{subequations}

\begin{align}
\tau_{1} & =\begin{cases}
\frac{6\mu}{\rho_{0}c_{s}^{2}} & viscous\thinspace fluids\\
\tau_{0}\left(\frac{\sigma_{0}}{\left\Vert \dev\left(\sigma\right)\right\Vert _{F}}\right)^{n} & elastoplastic\thinspace solids
\end{cases}\label{eq:tau1}\\
\tau_{2} & =\frac{\rho_{0}\kappa}{T_{0}c_{t}^{2}}
\end{align}

\end{subequations}

The justification of these choices is that classical Navier\textendash Stokes\textendash Fourier
theory is recovered in the stiff limit $\tau_{1},\tau_{2}\rightarrow0$
(see \citet{dumbser_high_2015}). The power law for elastoplastic
solids is based on material from \citet{barton_eulerian_2011}.

Finally, it is straightforward to verify that as a consequence of
\eqref{eq:p def}, \eqref{eq:sig def}, \eqref{eq:T def}, \eqref{eq:q def},
we have the following relations:

\begin{subequations}

\begin{align}
\sigma & =-\rho c_{s}^{2}G\dev\left(G\right)\\
\boldsymbol{q} & =c_{t}^{2}T\boldsymbol{J}\\
-\frac{\psi}{\theta_{1}(\tau_{1})} & =-\frac{3}{\tau_{1}}\left|A\right|^{\frac{5}{3}}A\dev\left(G\right)\\
-\frac{\rho\boldsymbol{H}}{\theta_{2}\left(\tau_{2}\right)} & =-\frac{T\rho_{0}}{T_{0}\tau_{2}}\boldsymbol{J}
\end{align}

\end{subequations}

The following constraint also holds (see \citet{peshkov_hyperbolic_2016}):

\begin{equation}
\det\left(A\right)=\frac{\rho}{\rho_{0}}
\end{equation}

The GPR model and Godunov and Romenski's 1970s model of elastoplastic
deformation in fact rely upon the same equations. The realization
of Peshkov and Romenski was that these are the equations of motion
for an arbitrary continuum - not just a solid - and so the model can
be applied to fluids too. Unlike in previous continuum models, material
elements have not only finite size, but also internal structure, encoded
in the distortion tensor.

The strain dissipation time $\tau_{1}$ of the GPR model is a continuous
analogue of Frenkel's ``particle settled life time'' (detailed in
\citet{frenkel_kinetic_1947}); the characteristic time taken for
a particle to move by a distance of the same order of magnitude as
the particle's size. Thus, $\tau_{1}$ characterizes the time taken
for a material element to rearrange with its neighbors. $\tau_{1}=\infty$
for solids and $\tau_{1}=0$ for inviscid fluids. It is in this way
that the GPR model seeks to describe all three major phases of matter,
as long as a continuum description is appropriate for the material
at hand.

The evolution equation for $\boldsymbol{J}$ and its contribution
to the energy of the system are derived from Romenski's model of hyperbolic
heat transfer, originally proposed in \citet{malyshev_hyperbolic_1986,romenski_hyperbolic_1989},
and implemented in \citet{romenski_conservative_2007,romenski_conservative_2010}.
In this model, $\boldsymbol{J}$ is effectively defined as the variable
conjugate to the entropy flux, in the sense that the latter is the
derivative of the specific internal energy with respect to $\boldsymbol{J}$.
Romenski remarks that it is more convenient to evolve $\boldsymbol{J}$
and $E$ than the heat flux or the entropy flux, and thus the equations
take the form given here. $\tau_{2}$ characterizes the speed of relaxation
of the thermal impulse due to heat exchange between material elements.

\section{Power-Law Fluids\label{sec:Power-Law-Fluids}}

\begin{figure*}
\begin{centering}
\includegraphics[height=0.2\paperheight]{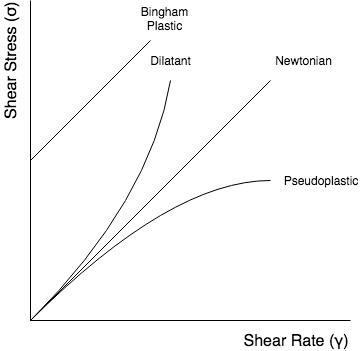}
\par\end{centering}
\caption{\label{fig:rheology}The stress-strain relationships for different
kinds of fluids}
\end{figure*}

The stress-strain relationships for various kinds of fluids are shown
in \prettyref{fig:rheology}. Dilatants and pseudoplastics may be
modelled using the following power law, with $n>1$ and $0<n<1$,
respectively:

\begin{align}
\boldsymbol{\sigma} & =K\left|\boldsymbol{\dot{\gamma}}\right|^{n-1}\boldsymbol{\dot{\gamma}}\\
\boldsymbol{\dot{\gamma}} & =\nabla\boldsymbol{v}+\nabla\boldsymbol{v}^{T}-\frac{2\tr\left(\nabla\boldsymbol{v}\right)}{3}I
\end{align}

$K>0$ is known as the \textit{consistency}, and $K\left|\boldsymbol{\dot{\gamma}}\right|^{n-1}$
is the \textit{apparent viscosity}. The norm is taken to be:

\begin{equation}
\left|X\right|=\sqrt{\frac{1}{2}X_{ij}X_{ij}}=\frac{\left\Vert X\right\Vert _{F}}{\sqrt{2}}
\end{equation}

In \citet{dumbser_high_2015} it was noted that when expressing the
state variables as an asymptotic expansion in the relaxation parameter
$\tau_{1}$, to first order we have:

\begin{equation}
\boldsymbol{\sigma}=\frac{1}{6}\tau_{1}\rho_{0}c_{s}^{2}\left(\nabla\boldsymbol{v}+\nabla\boldsymbol{v}^{T}-\frac{2}{3}\tr\left(\nabla\boldsymbol{v}\right)\boldsymbol{I}\right)\label{eq:AsymptoticRelationship}
\end{equation}

Thus, for a power law fluid, we require that:

\begin{equation}
\frac{1}{6}\tau_{1}\rho_{0}c_{s}^{2}=K\left|\boldsymbol{\dot{\gamma}}\right|^{n-1}
\end{equation}

Taking moduli of both sides of \eqref{eq:AsymptoticRelationship},
we also have:

\begin{equation}
\left|\boldsymbol{\sigma}\right|=\frac{1}{6}\tau_{1}\rho_{0}c_{s}^{2}\left|\boldsymbol{\dot{\gamma}}\right|
\end{equation}

Combining these two relationships, we obtain:

\begin{equation}
\tau_{1}=\frac{6K^{\frac{1}{n}}}{\rho_{0}c_{s}^{2}}\left|\frac{1}{\boldsymbol{\sigma}}\right|^{\frac{1-n}{n}}\coloneqq\tau_{0}\left|\frac{1}{\boldsymbol{\sigma}}\right|^{\frac{1-n}{n}}\label{eq:=0003C4-power-law}
\end{equation}

\section{Numerical Schemes\label{sec:Numerical-Schemes}}

Note that \eqref{eq:DensityEquation}, \eqref{eq:MomentumEquation},
\eqref{eq:DistortionEquation}, \eqref{eq:ThermalEquation}, \eqref{eq:EnergyEquation}
can be written in the following form:

\begin{equation}
\frac{\partial\boldsymbol{Q}}{\partial t}+\boldsymbol{\nabla}\cdot\boldsymbol{F}\left(\boldsymbol{Q}\right)+\boldsymbol{B}\left(\boldsymbol{Q}\right)\cdot\nabla\boldsymbol{Q}=\boldsymbol{S}\left(\boldsymbol{Q}\right)
\end{equation}

As described in \citet{toro_reimann_2009}, a viable way to solve
inhomogeneous systems of PDEs is to employ an operator splitting.
That is, the following subsystems are solved:

\begin{subequations}

\begin{align}
\frac{\partial\boldsymbol{Q}}{\partial t}+\boldsymbol{\nabla}\cdot\boldsymbol{F}\left(\boldsymbol{Q}\right)+\boldsymbol{B}\left(\boldsymbol{Q}\right)\cdot\nabla\boldsymbol{Q} & =\boldsymbol{0}\label{eq:HomogeneousSubsystem}\\
\frac{d\boldsymbol{Q}}{dt} & =\boldsymbol{S}\left(\boldsymbol{Q}\right)\label{eq:ODESubsystem}
\end{align}

\end{subequations}

The advantage of this approach is that specialized solvers can be
employed to compute the results of the different subsystems. Let $H^{\delta t},S^{\delta t}$
be the operators that take data $\boldsymbol{Q}\left(x,t\right)$
to $\boldsymbol{Q}\left(x,t+\delta t\right)$ under systems \eqref{eq:HomogeneousSubsystem}
and \eqref{eq:ODESubsystem} respectively. A second-order scheme (in
time) for solving the full set of PDEs over time step $\left[0,\Delta t\right]$
is obtained by calculating $\boldsymbol{Q_{\Delta t}}$ using a Strang
splitting:

\begin{equation}
\boldsymbol{Q_{\Delta t}}=S^{\frac{\Delta t}{2}}H^{\Delta t}S^{\frac{\Delta t}{2}}\boldsymbol{Q_{0}}
\end{equation}

In the scheme proposed here, the homogeneous subsystem will be solved
using a WENO reconstruction of the data, followed by a finite volume
update, and the temporal ODEs will be solved with appropriate ODE
solvers. It should be noted that there are other choices of solvers
for the homogeneous system that could have been made (e.g. see MUSCL,
SLIC, and WAF, among others in \citet{toro_reimann_2009}). The WENO
method was chosen due to the arbitrarily high-order spatial reconstructions
it is able to produce.

Noting that $\frac{d\rho}{dt}=0$ over the ODE time step, the operator
$S$ entails solving the following systems:

\begin{subequations}

\begin{align}
\frac{dA}{dt} & =\frac{-3}{\tau_{1}}\left|A\right|^{\frac{5}{3}}A\dev\left(G\right)\label{eq:DistortionODE}\\
\frac{d\boldsymbol{J}}{dt} & =-\frac{1}{\tau_{2}}\frac{T\rho_{0}}{T_{0}\rho}\boldsymbol{J}\label{eq:ThermalODE}
\end{align}

\end{subequations}

These systems can be solved concurrently with a stiff ODE solver.
The Jacobians of these two systems to be used in an ODE solver are
given in the appendix of \citet{jackson_fast_2017}. However, these
systems can also be solved separately, using the analytical results
presented in \prettyref{subsec:The-Thermal-Impulse-ODEs} and \prettyref{subsec:The-Distortion-ODEs},
under specific assumptions. The second-order Strang splitting is then:

\begin{equation}
\boldsymbol{Q_{\Delta t}}=D^{\frac{\Delta t}{2}}T^{\frac{\Delta t}{2}}H^{\Delta t}T^{\frac{\Delta t}{2}}D^{\frac{\Delta t}{2}}\boldsymbol{Q_{0}}
\end{equation}

where $D^{\delta t},T^{\delta t}$ are the operators solving the distortion
and thermal impulse ODEs respectively, over time step $\delta t$.
This allows us to bypass the relatively computationally costly process
of solving these systems numerically.

\subsection{The Homogeneous System}

A WENO reconstruction of the cell-averaged data is performed at the
start of the time step (as described in \citet{dumbser_ader-weno_2013}).
Focusing on a single cell $C_{i}$ at time $t_{n}$, we have $\boldsymbol{w^{n}}\left(\boldsymbol{x}\right)=\boldsymbol{w^{n}}_{p}\Psi_{p}\left(\boldsymbol{\chi}\left(\boldsymbol{x}\right)\right)$
in $C_{i}$ where $\Psi_{p}$ is a tensor product of basis functions
in each of the spatial dimensions. The flux in $C$ is approximated
by $\boldsymbol{F}\left(\boldsymbol{x}\right)\approx\boldsymbol{F}\left(\boldsymbol{w}_{p}\right)\Psi_{p}\left(\boldsymbol{\chi}\left(\boldsymbol{x}\right)\right)$.
$\boldsymbol{w}_{p}$ are stepped forwards half a time step using
the update formula:

\begin{align}
\frac{\boldsymbol{w_{p}^{n+\frac{1}{2}}}-\boldsymbol{w_{p}^{n}}}{\Delta t/2}= & -\boldsymbol{F}\left(\boldsymbol{w_{k}^{n}}\right)\cdot\nabla\Psi_{k}\left(\boldsymbol{\chi_{p}}\right)\\
 & -\boldsymbol{B}\left(\boldsymbol{w_{p}^{n}}\right)\cdot\left(\boldsymbol{w_{k}^{n}}\nabla\Psi_{k}\left(\boldsymbol{\chi_{p}}\right)\right)\nonumber 
\end{align}

i.e.

\begin{equation}
\boldsymbol{w_{p}^{n+\frac{1}{2}}}=\boldsymbol{w_{p}^{n}}-\frac{\Delta t}{2\Delta x}\left(\begin{array}{c}
\boldsymbol{F}\left(\boldsymbol{w_{k}^{n}}\right)\cdot\nabla\Psi_{k}\left(\boldsymbol{\chi_{p}}\right)\\
+\boldsymbol{B}\left(\boldsymbol{w_{p}^{n}}\right)\cdot\left(\boldsymbol{w_{k}^{n}}\nabla\Psi_{k}\left(\boldsymbol{\chi_{p}}\right)\right)
\end{array}\right)\label{eq:WENO half step}
\end{equation}

where $\boldsymbol{\chi_{p}}$ is the node corresponding to $\Psi_{p}$.
This evolution to the middle of the time step is similar to that used
in the second-order MUSCL and SLIC schemes (see \citet{toro_reimann_2009})
and, as with those schemes, it is integral to giving the method presented
here its second-order accuracy. 

Integrating \eqref{eq:HomogeneousSubsystem} over $C$ gives:

\begin{equation}
\boldsymbol{Q_{i}^{n+1}}=\boldsymbol{Q_{i}^{n}}-\Delta t_{n}\left(\boldsymbol{P_{i}^{n+\frac{1}{2}}}+\boldsymbol{D_{i}^{n+\frac{1}{2}}}\right)
\end{equation}

where

\begin{subequations}

\begin{align}
\boldsymbol{Q_{i}^{n}} & =\frac{1}{V}\int_{C}\boldsymbol{Q}\left(\boldsymbol{x},t_{n}\right)d\boldsymbol{x}\\
\boldsymbol{P_{i}^{n+\frac{1}{2}}} & =\frac{1}{V}\int_{C}\boldsymbol{B}\left(\boldsymbol{Q}\left(\boldsymbol{x},t_{n+\frac{1}{2}}\right)\right)\cdot\nabla\boldsymbol{Q}\left(\boldsymbol{x},t_{n+\frac{1}{2}}\right)d\boldsymbol{x}\\
\boldsymbol{D_{i}^{n+\frac{1}{2}}} & =\frac{1}{V}\varoint_{\partial C}\boldsymbol{\mathcal{D}}\left(\boldsymbol{Q^{-}}\left(\boldsymbol{s},t_{n+\frac{1}{2}}\right),\boldsymbol{Q^{+}}\left(\boldsymbol{s},t_{n+\frac{1}{2}}\right)\right)d\boldsymbol{s}
\end{align}

\end{subequations}

where $V$ is the volume of $C$ and $\boldsymbol{Q^{-},Q^{+}}$ are
the interior and exterior extrapolated states at the boundary of $C$,
respectively.

Note that \eqref{eq:HomogeneousSubsystem} can be rewritten as:

\begin{equation}
\frac{\partial\boldsymbol{Q}}{\partial t}+\boldsymbol{M}\left(\boldsymbol{Q}\right)\cdot\nabla\boldsymbol{Q}=\boldsymbol{0}
\end{equation}

where $\boldsymbol{M}=\frac{\partial\boldsymbol{F}}{\partial\boldsymbol{Q}}+\boldsymbol{B}$.
Let $\boldsymbol{n}$ be the normal to the boundary at point $\boldsymbol{s}\in\partial C$.
For the GPR model, $\hat{M}=\boldsymbol{M}\left(\boldsymbol{Q}\left(\boldsymbol{s}\right)\right)\cdot\boldsymbol{n}$
is a diagonalizable matrix with decomposition $\hat{M}=\hat{R}\hat{\Lambda}\hat{R}^{-1}$
where the columns of $\hat{R}$ are the right eigenvectors and $\hat{\Lambda}$
is the diagonal matrix of eigenvalues. Define also $\boldsymbol{\hat{F}}=\boldsymbol{F}\cdot\boldsymbol{n}$
and $\hat{B}=\boldsymbol{B}\cdot\boldsymbol{n}$. Using these definitions,
the interface terms arising in the FV formula have the following form:

\begin{align}
\boldsymbol{\mathcal{D}}\left(\boldsymbol{Q^{-}},\boldsymbol{Q^{+}}\right) & =\frac{1}{2}\left(\boldsymbol{\hat{F}}\left(\boldsymbol{Q^{+}}\right)+\boldsymbol{\hat{F}}\left(\boldsymbol{Q^{-}}\right)\right)\\
 & +\frac{1}{2}\left(+\tilde{B}\left(\boldsymbol{Q^{+}}-\boldsymbol{Q^{-}}\right)+\tilde{M}\left(\boldsymbol{Q^{+}}-\boldsymbol{Q^{-}}\right)\right)\nonumber 
\end{align}

$\tilde{M}$ is chosen to either correspond to a Rusanov/Lax-Friedrichs
flux (see \citet{toro_reimann_2009}):

\begin{equation}
\tilde{M}=\max\left(\max\left|\hat{\Lambda}\left(\boldsymbol{Q^{+}}\right)\right|,\max\left|\hat{\Lambda}\left(\boldsymbol{Q^{-}}\right)\right|\right)
\end{equation}

or a Roe flux (see \citet{dumbser_simple_2011}):
\begin{equation}
\hat{M}=\left|\int_{0}^{1}M\left(\boldsymbol{\boldsymbol{q^{-}}}+z\left(\boldsymbol{q^{+}}-\boldsymbol{q^{-}}\right)\right)dz\right|
\end{equation}

or a simplified Osher\textendash Solomon flux (see \citet{dumbser_simple_2011,dumbser_universal_2011}):

\begin{equation}
\tilde{M}=\int_{0}^{1}\left|\hat{M}\left(\boldsymbol{Q^{-}}+z\left(\boldsymbol{Q^{+}}-\boldsymbol{Q^{-}}\right)\right)\right|dz
\end{equation}

where

\begin{equation}
\left|\hat{M}\right|=\hat{R}\left|\hat{\Lambda}\right|\hat{R}^{-1}
\end{equation}

$\tilde{B}$ takes the following form:

\begin{equation}
\tilde{B}=\int_{0}^{1}\hat{B}\left(\boldsymbol{Q^{-}}+z\left(\boldsymbol{Q^{+}}-\boldsymbol{Q^{-}}\right)\right)dz
\end{equation}

$\boldsymbol{P_{i}^{n+\frac{1}{2}}},\boldsymbol{D_{i}^{n+\frac{1}{2}}}$
are calculated using an $N+1$-point Gauss-Legendre quadrature, replacing
$\boldsymbol{Q}\left(\boldsymbol{x},t_{n+\frac{1}{2}}\right)$ with
$\boldsymbol{w^{n+\frac{1}{2}}}\left(\boldsymbol{x}\right)$.

\subsection{The Thermal Impulse ODEs\label{subsec:The-Thermal-Impulse-ODEs}}

The following analytical solution to the thermal impulse ODEs was
first presented in \citet{jackson_fast_2017}. It is included here
for completeness.

Taking the EOS for the GPR model \eqref{eq:EnergyDefinition} and
denoting by $E_{2}^{\left(A\right)},E_{2}^{\left(J\right)}$ the components
of $E_{2}$ depending on $A$ and $\boldsymbol{J}$ respectively,
we have:

\begin{align}
T & =\frac{E_{1}}{c_{v}}\\
 & =\frac{E-E_{2}^{\left(A\right)}\left(\rho,s,A\right)-E_{3}\left(\boldsymbol{v}\right)}{c_{v}}-\frac{1}{c_{v}}E_{2}^{\left(J\right)}\left(\boldsymbol{J}\right)\nonumber \\
 & =c_{1}-c_{2}\left\Vert \boldsymbol{J}\right\Vert ^{2}\nonumber 
\end{align}

where:

\begin{subequations}

\begin{align}
c_{1} & =\frac{E-E_{2}^{\left(A\right)}\left(A\right)-E_{3}\left(\boldsymbol{v}\right)}{c_{v}}\\
c_{2} & =\frac{c_{t}^{2}}{2c_{v}}
\end{align}

\end{subequations}

Over the time period of the ODE \eqref{eq:ThermalODE}, $c_{1},c_{2}>0$
are constant. We have:

\begin{equation}
\frac{dJ_{i}}{dt}=-\left(\frac{1}{\tau_{2}}\frac{\rho_{0}}{T_{0}\rho}\right)J_{i}\left(c_{1}-c_{2}\left\Vert \boldsymbol{J}\right\Vert ^{2}\right)
\end{equation}

Therefore:

\begin{equation}
\frac{d}{dt}\left(J_{i}^{2}\right)=J_{i}^{2}\left(-a+b\left(J_{1}^{2}+J_{2}^{2}+J_{3}^{2}\right)\right)
\end{equation}

where

\begin{subequations}

\begin{align}
a & =\frac{2\rho_{0}}{\tau_{2}T_{0}\rho c_{v}}\left(E-E_{2}^{\left(A\right)}\left(A\right)-E_{3}\left(\boldsymbol{v}\right)\right)\\
b & =\frac{\rho_{0}c_{t}^{2}}{\tau_{2}T_{0}\rho c_{v}}
\end{align}

\end{subequations}

Note that this is a generalized Lotka-Volterra system in $\left\{ J_{1}^{2},J_{2}^{2},J_{3}^{2}\right\} $.
It has the following analytical solution:

\begin{equation}
\boldsymbol{J}\left(t\right)=\boldsymbol{J}\left(0\right)\sqrt{\frac{1}{e^{at}-\frac{b}{a}\left(e^{at}-1\right)\left\Vert \boldsymbol{J}\left(0\right)\right\Vert ^{2}}}
\end{equation}

\subsection{The Distortion ODEs\label{subsec:The-Distortion-ODEs}}

The following analytical solution to the distortion ODEs for Newtonian
fluids was first presented in \citet{jackson_fast_2017}. It is included
here, as the solutions for non-Newtonian fluids and elastoplastic
solids depend on the Newtonian solution.

\subsubsection{Newontian Fluids\label{subsec:Analytical-Approximation}}

Let $k_{0}=\frac{3}{\tau_{1}}\left(\frac{\rho}{\rho_{0}}\right)^{\frac{5}{3}}>0$
and let $A$ have singular value decomposition $U\Sigma V^{T}$. Then:

\begin{equation}
G=\left(U\Sigma V^{T}\right)^{T}U\Sigma V^{T}=V\Sigma^{2}V^{T}
\end{equation}

\begin{equation}
\tr\left(G\right)=\tr\left(V\Sigma^{2}V^{T}\right)=\tr\left(\Sigma^{2}V^{T}V\right)=\tr\left(\Sigma^{2}\right)
\end{equation}

Therefore:

\begin{align}
\frac{dA}{dt} & =-k_{0}U\Sigma V^{T}\left(V\Sigma^{2}V^{T}-\frac{\tr\left(\Sigma^{2}\right)}{3}I\right)\\
 & =-k_{0}U\Sigma\left(\Sigma^{2}-\frac{\tr\left(\Sigma^{2}\right)}{3}\right)V^{T}\nonumber \\
 & =-k_{0}U\Sigma\dev\left(\Sigma^{2}\right)V^{T}\nonumber 
\end{align}

It is a common result (see \citet{giles_extended_2008}) that:

\begin{equation}
d\Sigma=U^{T}dAV
\end{equation}

and thus:

\begin{equation}
\frac{d\Sigma}{dt}=-k_{0}\Sigma\dev\left(\Sigma^{2}\right)
\end{equation}

Using a fast $3\times3$ SVD algorithm (such as in \citet{mcadams_computing_2011}),
$U,V,\Sigma$ can be obtained, after which the following procedure
is applied to $\Sigma$, giving $A\left(t\right)=U\Sigma\left(t\right)V^{T}$.

Denote the singular values of $A$ by $a_{1},a_{2},a_{3}$. Then:

{\small{}
\begin{equation}
\Sigma\dev\left(\Sigma^{2}\right)=\left(\begin{array}{ccc}
a_{1}\left(a_{1}^{2}-\alpha\right) & 0 & 0\\
0 & a_{1}\left(a_{1}^{2}-\alpha\right) & 0\\
0 & 0 & a_{1}\left(a_{1}^{2}-\alpha\right)
\end{array}\right)
\end{equation}
}{\small\par}

where

\begin{equation}
\alpha=\frac{a_{1}^{2}+a_{2}^{2}+a_{3}^{2}}{3}
\end{equation}

Letting $x_{i}=\frac{a_{i}^{2}}{\det\left(A\right)^{\frac{2}{3}}}=\frac{a_{i}^{2}}{\left(\frac{\rho}{\rho_{0}}\right)^{\frac{2}{3}}}$
we have:

\begin{equation}
\frac{dx_{i}}{d\tau}=-3x_{i}\left(x_{i}-\bar{x}\right)\label{eq:StretchODESystem}
\end{equation}

where $\tau=\frac{2}{\tau_{1}}\left(\frac{\rho}{\rho_{0}}\right)^{\frac{7}{3}}t$
and $\bar{x}$ is the arithmetic mean of $x_{1},x_{2},x_{3}$. This
ODE system travels along the surface $\Psi=\left\{ x_{1},x_{2},x_{3}>0,x_{1}x_{2}x_{3}=1\right\} $
to the point $x_{1},x_{2},x_{3}=1$. This surface is symmetrical in
the planes $x_{1}=x_{2}$, $x_{1}=x_{3}$, $x_{2}=x_{3}$. As such,
given that the system is autonomous, the paths of evolution of the
$x_{i}$ cannot cross the intersections of these planes with $\Psi$.
Thus, any non-strict inequality of the form $x_{i}\geq x_{j}\geq x_{k}$
is maintained for the whole history of the system. By considering
\eqref{eq:StretchODESystem} it is clear that in this case $x_{i}$
is monotone decreasing, $x_{k}$ is monotone increasing, and the time
derivative of $x_{j}$ may switch sign.

We now explore cases when even the reduced ODE system \eqref{eq:StretchODESystem}
need not be solved numerically. Define the following variables:

\begin{subequations}

\begin{align}
m & =\frac{x_{1}+x_{2}+x_{3}}{3}\\
u & =\frac{\left(x_{1}-x_{2}\right)^{2}+\left(x_{2}-x_{3}\right)^{2}+\left(x_{3}-x_{1}\right)^{2}}{3}
\end{align}

\end{subequations}

It is a standard result that $m\geq\sqrt[3]{x_{1}x_{2}x_{3}}$. Thus,
$m\geq1$. Note that $u$ is proportional to the internal energy contribution
from the distortion. From \eqref{eq:StretchODESystem} we have:

\begin{subequations}

\begin{align}
\frac{du}{d\tau} & =-18\left(1-m\left(m^{2}-\frac{5}{6}u\right)\right)\\
\frac{dm}{d\tau} & =-u
\end{align}

\end{subequations}

Combining these equations, we have:

\begin{equation}
\frac{d^{2}m}{d\tau^{2}}=-\frac{du}{d\tau}=18\left(1-m\left(m^{2}-\frac{5}{6}u\right)\right)
\end{equation}

Therefore:

\begin{equation}
\left\{ \begin{array}{c}
\frac{d^{2}m}{d\tau^{2}}+15m\frac{dm}{d\tau}+18\left(m^{3}-1\right)=0\\
m\left(0\right)=m_{0}\\
m^{'}\left(0\right)=-u_{0}
\end{array}\right.
\end{equation}

We make the following assumption, noting that it is true in all physical
situations tested in this study:

\begin{equation}
m\left(t\right)=1+\eta\left(t\right),\quad\eta\ll1\;\forall t\geq0\label{eq:Assumption}
\end{equation}

Thus, we have the linearized ODE:

\begin{equation}
\left\{ \begin{array}{c}
\frac{d^{2}\eta}{d\tau^{2}}+15\frac{d\eta}{d\tau}+54\eta=0\\
\eta\left(0\right)=m_{0}-1\\
\eta^{'}\left(0\right)=-u_{0}
\end{array}\right.\label{eq:Linearized}
\end{equation}

This is a Sturm-Liouville equation with solution:

\begin{equation}
\eta\left(\tau\right)=\frac{e^{-9\tau}}{3}\left(ae^{3\tau}-b\right)
\end{equation}

where

\begin{subequations}

\begin{align}
a & =9m_{0}-u_{0}-9\\
b & =6m_{0}-u_{0}-6
\end{align}

\end{subequations}

Thus, we also have:

\begin{equation}
u\left(\tau\right)=e^{-9\tau}\left(2ae^{3\tau}-3b\right)
\end{equation}

Denote the following:

\begin{subequations}

\begin{align}
m_{\Delta t} & =1+\eta\left(\frac{2}{\tau_{1}}\left(\frac{\rho}{\rho_{0}}\right)^{\frac{7}{3}}\Delta t\right)\\
u_{\Delta t} & =u\left(\frac{2}{\tau_{1}}\left(\frac{\rho}{\rho_{0}}\right)^{\frac{7}{3}}\Delta t\right)
\end{align}

\end{subequations}

Once these have been found, we have:

\begin{subequations}

\begin{align}
\frac{x_{i}+x_{j}+x_{k}}{3} & =m_{\Delta t}\\
\frac{\left(x_{i}-x_{j}\right)^{2}+\left(x_{j}-x_{k}\right)^{2}+\left(x_{k}-x_{i}\right)^{2}}{3} & =u_{\Delta t}\\
x_{i}x_{j}x_{k} & =1
\end{align}

\end{subequations}

This gives:

\begin{subequations}

\begin{align}
x_{i} & =\frac{\Xi}{6}+\frac{u_{\Delta t}}{\Xi}+m_{\Delta t}\\
x_{j} & =\frac{1}{2}\left(\sqrt{\frac{x_{i}\left(3m_{\Delta t}-x_{i}\right)^{2}-4}{x_{i}}}+3m_{\Delta t}-x_{i}\right)\label{eq:xj}\\
x_{k} & =\frac{1}{x_{i}x_{j}}\label{eq:xk}
\end{align}

\end{subequations}

where

\begin{subequations}

\begin{align}
\Xi & =\sqrt[3]{6\left(\sqrt{81\Delta^{2}-6u_{\Delta t}^{3}}+9\Delta\right)}\\
\Delta & =-2m_{\Delta t}^{3}+m_{\Delta t}u_{\Delta t}+2
\end{align}

\end{subequations}

Note that taking the real parts of the above expression for $x_{i}$
gives:

\begin{subequations}

\begin{align}
x_{i} & =\frac{\sqrt{6u_{\Delta t}}}{3}\cos\left(\frac{\theta}{3}\right)+m_{\Delta t}\label{eq:xi}\\
\theta & =\tan^{-1}\left(\frac{\sqrt{6u_{\Delta t}^{3}-81\Delta^{2}}}{9\Delta}\right)\label{eq:=0003B8}
\end{align}

\end{subequations}

At this point it is not clear which values of $\left\{ x_{i},x_{j},x_{k}\right\} $
are taken by $x_{1},x_{2},x_{3}$. However, this can be inferred from
the fact that any relation $x_{i}\geq x_{j}\geq x_{k}$ is maintained
over the lifetime of the system. Thus, the stiff ODE solver has been
obviated by a few arithmetic operations.

\subsubsection{Power Law Fluids}

Take the singular value decomposition $A=U\Sigma V^{T}$. Note that:

\begin{equation}
\sigma=-\rho c_{s}^{2}A^{T}A\dev\left(A^{T}A\right)=-\rho c_{s}^{2}V\Sigma^{2}\dev\left(\Sigma^{2}\right)V^{T}
\end{equation}

Thus:

\begin{equation}
\left\Vert \sigma\right\Vert _{F}^{k}=\rho^{k}c_{s}^{2k}\left\Vert \Sigma^{2}\dev\left(\Sigma^{2}\right)\right\Vert _{F}^{k}
\end{equation}

Thus, according to \prettyref{eq:=0003C4-power-law}, and letting
$k=\frac{1-n}{n}$, we have:

\begin{equation}
\frac{d\Sigma}{dt}=-\frac{3}{\tau_{0}}\left(\frac{\rho}{\rho_{0}}\right)^{\frac{5}{3}}\frac{\rho^{k}c_{s}^{2k}}{2^{\frac{k}{2}}}\left\Vert \Sigma^{2}\dev\left(\Sigma^{2}\right)\right\Vert _{F}^{k}\Sigma\dev\left(\Sigma^{2}\right)
\end{equation}

Letting $x_{i}=\frac{a_{i}^{2}}{\det\left(A\right)^{\frac{2}{3}}}=\frac{a_{i}^{2}}{\left(\frac{\rho}{\rho_{0}}\right)^{\frac{2}{3}}}$
then $\Sigma^{2}=\det\left(A\right)^{\frac{2}{3}}X$ where $X=\diag\left(x_{1},x_{2},x_{3}\right)$.
Thus, we have:

\begin{equation}
\frac{dx_{i}}{d\tilde{t}}=-3\left\Vert X\dev\left(X\right)\right\Vert _{F}^{k}x_{i}\left(x_{i}-\bar{x}\right)\label{eq:StretchODESystem-2-1}
\end{equation}

where:

\begin{equation}
\tilde{t}=\frac{2}{\tau_{0}}\left(\frac{\rho}{\rho_{0}}\right)^{\frac{4k+7}{3}}\left(\frac{\rho c_{s}^{2}}{\sqrt{2}}\right)^{k}t
\end{equation}

Note that:

\begin{align}
9\left\Vert X\dev\left(X\right)\right\Vert _{F}^{2} & =4\left(x_{1}^{4}+x_{2}^{4}+x_{3}^{4}\right)\\
 & -2\left(x_{1}^{2}x_{2}^{2}+x_{3}^{2}x_{2}^{2}+x_{1}^{2}x_{3}^{2}\right)\nonumber \\
 & +\sum_{i\neq j,j\neq k,k\neq i}x_{i}^{2}x_{j}x_{k}-4\sum_{i\neq j}x_{i}^{3}x_{j}\nonumber 
\end{align}

Defining $m,u$ as before, we have:

\begin{equation}
\left\Vert X\dev\left(X\right)\right\Vert _{F}^{2}=\frac{1}{2}u^{2}+4m^{2}u-6m^{4}+6m
\end{equation}

This leads to the following coupled system of ODEs:

\begin{subequations}

\begin{align}
\frac{du}{d\tilde{t}} & =-18\frac{d\tau}{d\tilde{t}}\left(1-m\left(m^{2}-\frac{5}{6}u\right)\right)\\
\frac{dm}{d\tilde{t}} & =-\frac{d\tau}{d\tilde{t}}u
\end{align}

\end{subequations}

where we have defined the variable $\tau$ by:

\begin{equation}
\frac{d\tau}{d\tilde{t}}=\left(\frac{1}{2}u^{2}+4m^{2}u-6m^{4}+6m\right)^{\frac{k}{2}}
\end{equation}

Using the approximation solution from before:

\begin{subequations}

\begin{align}
m\left(\tau\right) & =1+\frac{e^{-9\tau}}{3}\left(ae^{3\tau}-b\right)\label{eq:m(=0003C4)-1}\\
u\left(\tau\right) & =e^{-9\tau}\left(2ae^{3\tau}-3b\right)\label{eq:u(=0003C4)-1}
\end{align}

\end{subequations}

It is straightforward to verify that:

\begin{align}
\frac{d\tau}{d\tilde{t}} & =\frac{1}{54^{\frac{k}{2}}}\left(\begin{array}{c}
108ae^{-6\tau}-324be^{-9\tau}\\
+180a^{2}e^{-12\tau}-612abe^{-15\tau}\\
+459b^{2}e^{-18\tau}-24a^{2}be^{-21\tau}\\
+\left(48ab^{2}-4a^{4}\right)e^{-24\tau}\\
+\left(16a^{3}b-24b^{3}\right)e^{-27\tau}\\
-24a^{2}b^{2}e^{-30\tau}+16ab^{3}e^{-33\tau}\\
-4b^{4}e^{-36\tau}
\end{array}\right)^{\frac{k}{2}}\\
 & \equiv\frac{f\left(\tau\right)^{\frac{k}{2}}}{54^{\frac{k}{2}}}\nonumber 
\end{align}

$f\left(\tau\right)$ is approximated by $g\left(\tau\right)\equiv ce^{-\frac{c}{\lambda}\tau}$,
where:

\begin{subequations}

\begin{align}
c & =108a-324b+180a^{2}-612ab+459b^{2}\\
 & -24\left(a^{2}b-2ab^{2}+b^{3}\right)-4\left(a-b\right)^{4}\nonumber \\
\lambda & =18a-36b+15a^{2}-\frac{204ab}{5}+\frac{51b^{2}}{2}\\
 & -\frac{8a^{2}b}{7}+2ab^{2}-\frac{8b^{3}}{9}-\frac{a^{4}}{6}+\frac{16a^{3}b}{27}\nonumber \\
 & -\frac{4a^{2}b^{2}}{5}+\frac{16ab^{3}}{33}-\frac{b^{4}}{9}\nonumber 
\end{align}

\end{subequations}

Note that $f\left(0\right)=g\left(0\right)$ and $\int_{0}^{\infty}\left(f\left(\tau\right)-g\left(\tau\right)\right)d\tau=0$.
Thus, we have:

\begin{equation}
\frac{d\tau}{d\tilde{t}}\approx\left(\frac{c}{54}\right)^{\frac{k}{2}}e^{-\frac{kc}{2\lambda}\tau}
\end{equation}

Therefore:

\begin{align}
\tau & \approx\frac{2\lambda}{kc}\log\left(\frac{kc}{2\lambda}\left(\frac{c}{54}\right)^{\frac{k}{2}}\tilde{t}+1\right)\\
 & =\frac{2\lambda}{kc}\log\left(\frac{kc}{\tau_{0}\lambda}\left(\frac{\rho}{\rho_{0}}\right)^{\frac{4k+7}{3}}\left(\frac{\sqrt{c}\rho c_{s}^{2}}{6\sqrt{3}}\right)^{k}t+1\right)\nonumber 
\end{align}

\subsubsection{Elastoplastic Solids}

For elastoplastic materials governed by the power law described in
\eqref{eq:tau1}:

\begin{equation}
\frac{d\Sigma}{dt}=-\frac{3}{\tau_{0}}\left(\frac{\rho}{\rho_{0}}\right)^{\frac{5}{3}}\frac{\left(\frac{3}{2}\right)^{\frac{n}{2}}\rho^{n}c_{s}^{2n}\left\Vert \dev\left(\Sigma^{2}\dev\left(\Sigma^{2}\right)\right)\right\Vert _{F}^{n}}{\sigma_{0}^{n}}\Sigma\dev\left(\Sigma^{2}\right)
\end{equation}

Thus, we have:

\begin{equation}
\frac{dx_{i}}{d\tilde{t}}=-3\left\Vert \dev\left(X\dev\left(X\right)\right)\right\Vert _{F}^{n}x_{i}\left(x_{i}-\bar{x}\right)\label{eq:StretchODESystem-2}
\end{equation}

where:

\begin{equation}
\tilde{t}=\frac{2}{\tau_{0}}\left(\frac{\rho}{\rho_{0}}\right)^{\frac{4n+7}{3}}\left(\sqrt{\frac{3}{2}}\frac{\rho c_{s}^{2}}{\sigma_{0}}\right)^{n}t
\end{equation}

Note that:

\begin{align}
\frac{27}{2}\left\Vert \dev\left(X\dev\left(X\right)\right)\right\Vert _{F}^{2} & =\frac{3}{2}\sum_{i\neq j,j\neq k,k\neq i}x_{i}^{2}x_{j}x_{k}\\
 & -2\sum_{i\neq j}x_{i}^{3}x_{j}\nonumber \\
 & -3\left(x_{1}^{2}x_{2}^{2}+x_{3}^{2}x_{2}^{2}+x_{1}^{2}x_{3}^{2}\right)\nonumber \\
 & +4\left(x_{1}^{4}+x_{2}^{4}+x_{3}^{4}\right)\nonumber 
\end{align}

Thus we have:

\begin{equation}
\left\Vert \dev\left(X\dev\left(X\right)\right)\right\Vert _{F}^{2}=\frac{1}{6}u^{2}+4m^{2}u-6m^{4}+6m
\end{equation}

This leads to the following coupled system of ODEs:

\begin{subequations}

\begin{align}
\frac{du}{d\tilde{t}} & =-18\frac{d\tau}{d\tilde{t}}\left(1-m\left(m^{2}-\frac{5}{6}u\right)\right)\\
\frac{dm}{d\tilde{t}} & =-\frac{d\tau}{d\tilde{t}}u
\end{align}

\end{subequations}

where we have defined the variable $\tau$ by:

\begin{equation}
\frac{d\tau}{d\tilde{t}}=\left(\frac{1}{6}u^{2}+4m^{2}u-6m^{4}+6m\right)^{\frac{n}{2}}
\end{equation}

Then we have:

\begin{subequations}

\begin{align}
\frac{du}{d\tau} & =-18\left(1-m\left(m^{2}-\frac{5}{6}u\right)\right)\\
\frac{dm}{d\tau} & =-u
\end{align}

\end{subequations}

Using the approximate solution \eqref{eq:m(=0003C4)-1}, \eqref{eq:u(=0003C4)-1}
again, it is straightforward to verify that:

\begin{align}
\frac{d\tau}{d\tilde{t}} & =\frac{1}{54^{\frac{n}{2}}}\left(\begin{array}{c}
108ae^{-6\tau}-324be^{-9\tau}\\
+108a^{2}e^{-12\tau}-396abe^{-15\tau}\\
+297b^{2}e^{-18\tau}-24a^{2}be^{-21\tau}\\
+\left(48ab^{2}-4a^{4}\right)e^{-24\tau}\\
+\left(16a^{3}b-24b^{3}\right)e^{-27\tau}\\
-24a^{2}b^{2}e^{-30\tau}+16ab^{3}e^{-33\tau}\\
-4b^{4}e^{-36\tau}
\end{array}\right)^{\frac{n}{2}}\\
 & \equiv\frac{f\left(\tau\right)^{\frac{n}{2}}}{54^{\frac{n}{2}}}\nonumber 
\end{align}

$f\left(\tau\right)$ is approximated by $g\left(\tau\right)\equiv ce^{-\lambda\tau}$,
where:

\begin{subequations}

\begin{align}
c & =108a-324b+108a^{2}-396ab+297b^{2}\\
 & -24\left(a^{2}b-2ab^{2}+b^{3}\right)-4\left(a-b\right)^{4}\nonumber \\
\lambda & =18a-36b+9a^{2}-\frac{132ab}{5}+\frac{33b^{2}}{2}\\
 & -\frac{8a^{2}b}{7}+2ab^{2}-\frac{8b^{3}}{9}-\frac{a^{4}}{6}\nonumber \\
 & +\frac{16a^{3}b}{27}-\frac{4a^{2}b^{2}}{5}+\frac{16ab^{3}}{33}-\frac{b^{4}}{9}\nonumber 
\end{align}

\end{subequations}

Note that $f\left(0\right)=g\left(0\right)$ and $\int_{0}^{\infty}\left(f\left(\tau\right)-g\left(\tau\right)\right)d\tau=0$.
Thus, we have:

\begin{equation}
\frac{d\tau}{d\tilde{t}}\approx\left(\frac{c}{54}\right)^{\frac{n}{2}}e^{-\frac{nc}{2\lambda}\tau}
\end{equation}

Therefore:

\begin{align}
\tau & \approx\frac{2\lambda}{nc}\log\left(\frac{nc}{2\lambda}\left(\frac{c}{54}\right)^{\frac{n}{2}}\tilde{t}+1\right)\\
 & =\frac{2\lambda}{nc}\log\left(\frac{nc}{\tau_{0}\lambda}\left(\frac{\rho}{\rho_{0}}\right)^{\frac{4n+7}{3}}\left(\frac{\sqrt{c}}{6}\frac{\rho c_{s}^{2}}{\sigma_{0}}\right)^{n}t+1\right)\nonumber 
\end{align}

Thus, the value of $A$ at time $\Delta t$ is found by substituting
the following into \eqref{eq:m(=0003C4)-1}, \eqref{eq:u(=0003C4)-1}:

\begin{equation}
\tau=\frac{2\lambda}{nc}\log\left(\frac{nc}{\tau_{0}\lambda}\left(\frac{\rho}{\rho_{0}}\right)^{\frac{4n+7}{3}}\left(\frac{\sqrt{c}}{6}\frac{\rho c_{s}^{2}}{\sigma_{0}}\right)^{n}\Delta t+1\right)
\end{equation}

The results are in turn substituted into \eqref{eq:xi}, \eqref{eq:xj},
\eqref{eq:xk}.

\subsection{Distortion Correction in Fluids}

Owing to the linearization step in \eqref{eq:Linearized}, the method
presented will perform poorly if the mean of the normalized singular
values of the distortion tensor, $m$, deviates significantly from
$1$. To avert this, the following resetting procedure was applied
globally for fluid flow problems when $m>1.03$:

\begin{subequations}

\begin{align}
E & \mapsto E-\frac{c_{S}^{2}}{4}\left\Vert \dev\left(G\right)\right\Vert _{F}^{2}\\
A & \mapsto\left(\frac{\rho}{\rho_{0}}\right)^{1/3}I
\end{align}

\end{subequations}

This is justified by the fact that the distortion tensor is not a
macroscopically-measurable quantity. This transformation leaves the
density, pressure, and velocity of the fluid unchanged, and was found
to improve the stability of the numerical scheme, while at the same
time producing correct results, as demonstrated in the following section.

\section{Numerical Results\label{sec:Numerical-Results}}

In this section, a variety of test problems are solved, with a dual
purpose. Firstly, we demonstrate the ability of the modified GPR formulation
presented in \prettyref{sec:Power-Law-Fluids} to model power-law
fluids. Secondly, we demonstrate the efficacy of the numerical schemes
presented in \prettyref{sec:Numerical-Schemes} in solving this system,
and the existing power-law plasticity formulation of the GPR model.

\subsection{Strain Relaxation Test}

The aim of this test is to gauge the accuracy of the approximate analytic
solver for the distortion equations.

Take initial data used by Barton:

\begin{equation}
A=\left(\begin{array}{ccc}
1 & 0 & 0\\
-0.01 & 0.95 & 0.02\\
-0.015 & 0 & 0.9
\end{array}\right)^{-1}
\end{equation}

The following parameter values were used: $\rho_{0}=1,c_{s}=0.219,n=4,\sigma_{0}=9\times10^{-4},\tau_{0}=0.1$.

\begin{figure}[p]
\begin{centering}
\includegraphics[width=0.5\textwidth]{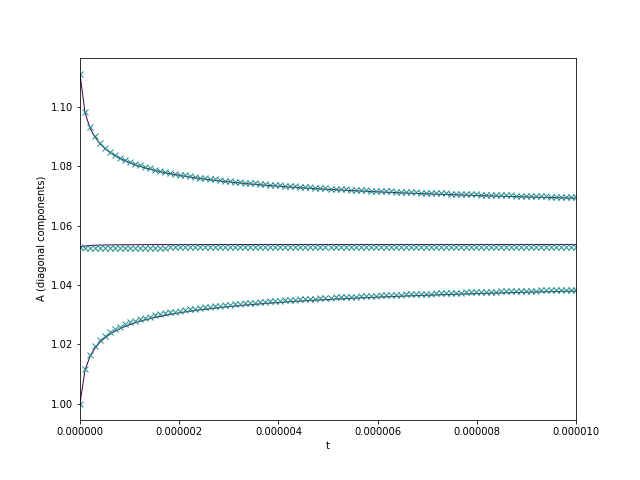}\includegraphics[width=0.5\textwidth]{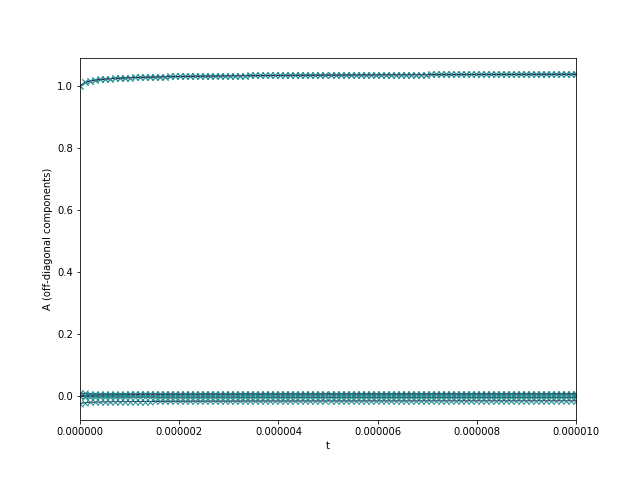}
\par\end{centering}
\caption{\label{fig:Distortion-ODEs_A-plastic}Distortion tensor components
during the Strain Relaxation Test: approximate analytical solution
(crosses) and numerical ODE solution (solid line)}
\end{figure}

\begin{figure}[p]
\begin{centering}
\includegraphics[width=0.5\textwidth]{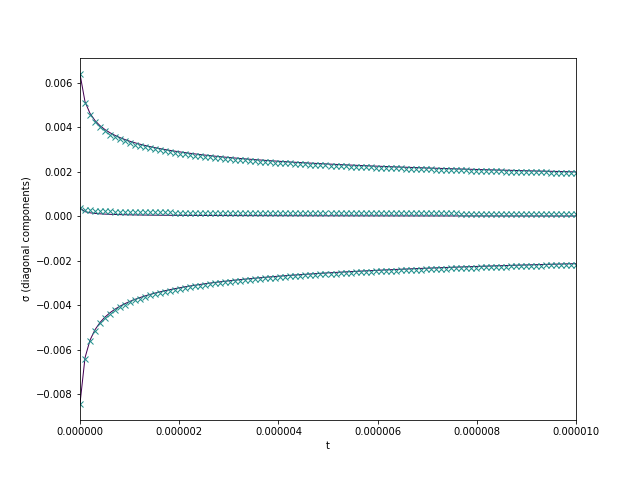}\includegraphics[width=0.5\textwidth]{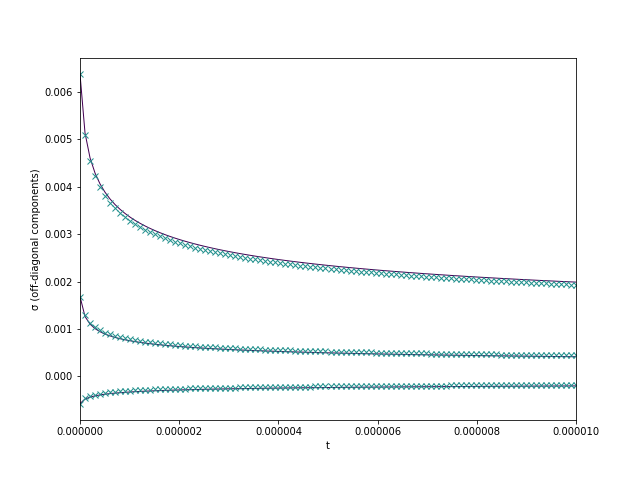}
\par\end{centering}
\caption{\label{fig:Distortion-ODEs_=0003C3-plastic}Stress tensor components
during the Strain Relaxation Test: approximate analytical solution
(crosses) and numerical ODE solution (solid line)}
\end{figure}

\begin{figure}[p]
\begin{centering}
\includegraphics[width=0.5\textwidth]{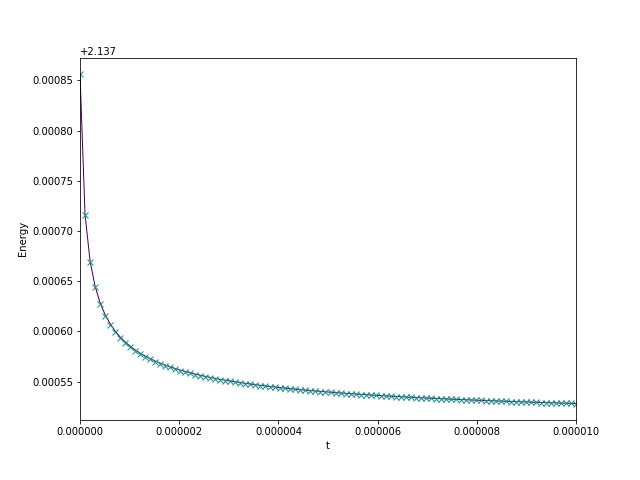}
\par\end{centering}
\caption{\label{fig:Distortion-ODEs_=000395-plastic}Total energy during the
Strain Relaxation Test: approximate analytical solution (crosses)
and numerical ODE solution (solid line)}
\end{figure}

The evolution of the components of the distortion tensor, according
to both the approximate analytical solver and a stiff numerical ODE
solver, are given in \prettyref{fig:Distortion-ODEs_A-plastic}, \prettyref{fig:Distortion-ODEs_=0003C3-plastic},
and \prettyref{fig:Distortion-ODEs_=000395-plastic}. As can be seen,
the approximate analytic solver compares well with the exact solution
for the distortion tensor $A$, and thus also the stress tensor and
the energy.

\subsection{Poiseuille Flow}

The aim of this test is to gauge both the performance of the modified
formulation of the GPR model in simulating power-law fluids, and the
accuracy of the new numerical scheme we have presented to solve it.
The problem of poiseuille flow has been chosen due to the availability
of an analytical solution against which to compare.

This test consists of a fluid traveling down a channel of constant
width $L$, with a constant pressure gradient $\Delta p$ along the
length of the channel. No-slip boundary conditions are imposed on
the channel walls. For a non-Newtonian fluid obeying a power law,
the steady-state velocity profile across the channel is given by \citet{ferras_analytical_2012}:

\begin{subequations}

\begin{align}
v & =\frac{\rho}{k}\left(\frac{\Delta p}{K}\right)^{1/n}\left(\left(\frac{L}{2}\right)^{k}-\left(x-\frac{L}{2}\right)^{k}\right)\\
k & =\frac{n+1}{n}
\end{align}

\end{subequations}

where $x\in\left[0,L\right]$.

In this case, $L=0.25$, $\Delta p=0.48$, $K=10^{-2}$. The fluid
is initially at rest, with $\rho_{0}=1$, $A=I$, $p=100/\gamma$.
It follows an ideal gas EOS with $\gamma=1.4$, $c_{s}=1$. The pressure
gradient is imposed by means of a body force, implemented as a constant
source term to the momentum equation.

The final time was taken to be $20$, so that in each case the system
had reached steady state. 100 cells were taken across the width of
the channel. A third order WENO method was used, with a CFL number
of 0.6.

\begin{figure*}
\begin{centering}
\includegraphics[width=0.5\textwidth]{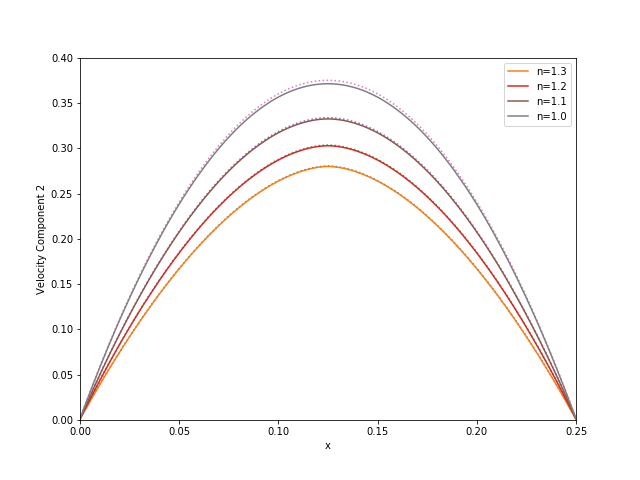}\includegraphics[width=0.5\textwidth]{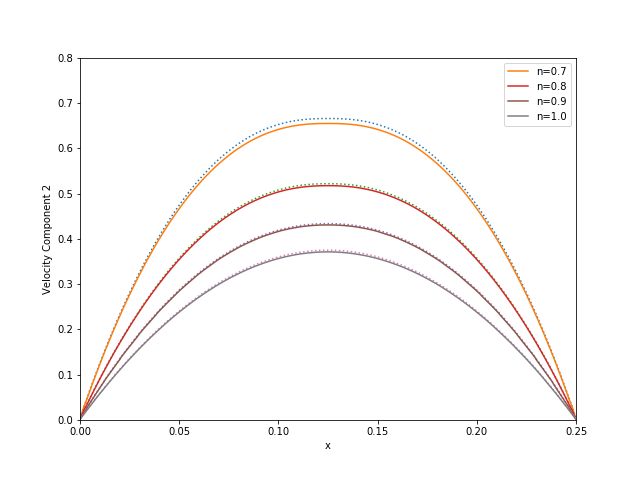}
\par\end{centering}
\caption{\label{fig:poiseuille}Velocity profiles for different dilatants (left)
and pseudoplastics (right), in steady Poiseuille flow}
\end{figure*}

Results for various values of $n$ are shown in \prettyref{fig:poiseuille}.
The exact solutions are shown as dotted lines, with the numerical
solutions in solid colors. Note that there is good agreement between
the numerical solutions and exact solutions for all values of $n$.

\subsection{Lid-Driven Cavity}

This lid-driven cavity test has been chosen here as a famous multidimensional
problem against which the power-law fluid framework we have presented
can be benchmarked. See \citet{sverdrup_highly_2018} for detailed
analysis of this problem, under power-law fluids and other non-Newtonian
fluids.

The test consists of a square grid, with one side at a constant velocity
of $1$, and the other three stationary, with no-slip boundary conditions
imposed. The fluid obeys an ideal gas EOS with $\gamma=1.4$ and $c_{s}=1$.
It obeys a viscosity power law with $K=10^{-2}$, for various $n$.
It is initially at rest, with $\rho=1$, $p=1$, $A=I$.

The grid is chosen to have size $100\times100$. A third order WENO
method is used, with a CFL number of 0.5.

\begin{figure*}
\begin{centering}
\includegraphics[width=0.5\textwidth]{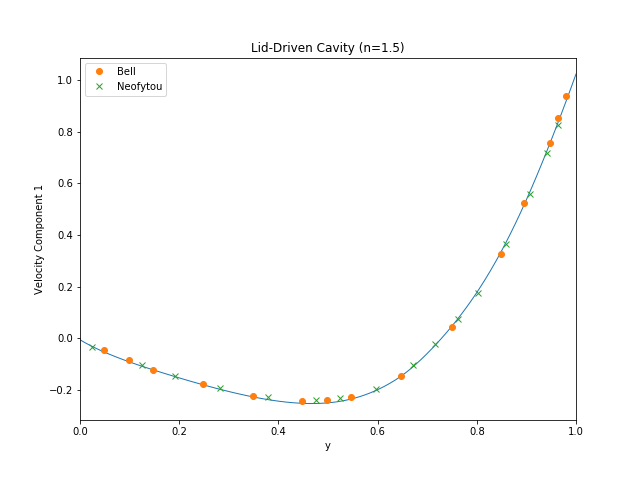}\includegraphics[width=0.5\textwidth]{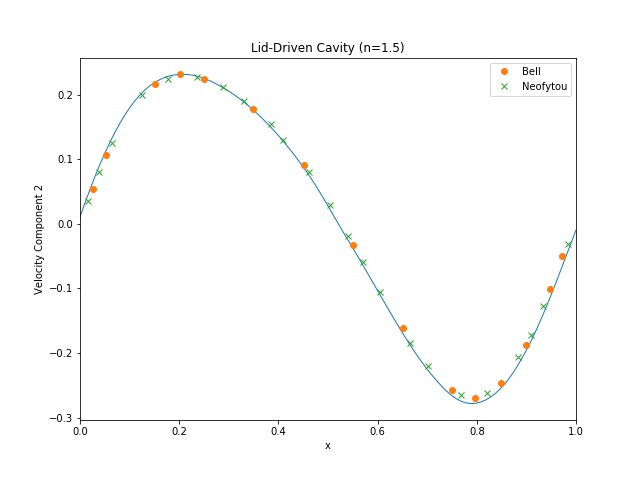}
\par\end{centering}
\caption{\label{fig:lid-driven-cavity-n=00003D1.5}Velocity profiles for the
Lid-Driven Cavity Test under our new formulation (solid line), for
a dilatant with n=1.5. Slices are taken through the center of the
domain, in both axes, and compared with those of \citet{bell_p-version_1994}
and \citet{neofytou_3rd_2005}.}
\end{figure*}

\begin{figure*}
\begin{centering}
\includegraphics[width=0.5\textwidth]{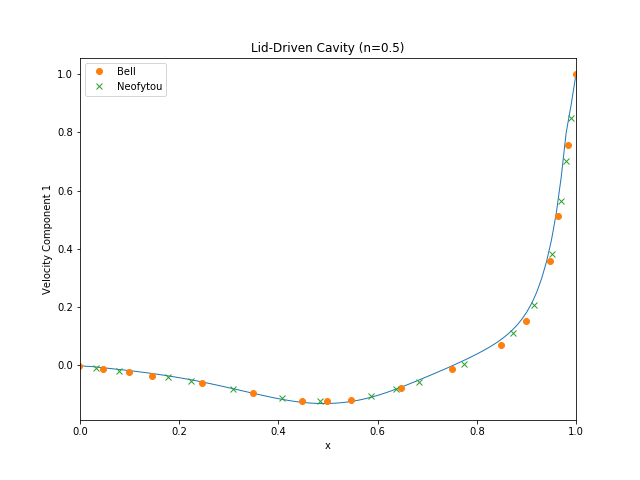}\includegraphics[width=0.5\textwidth]{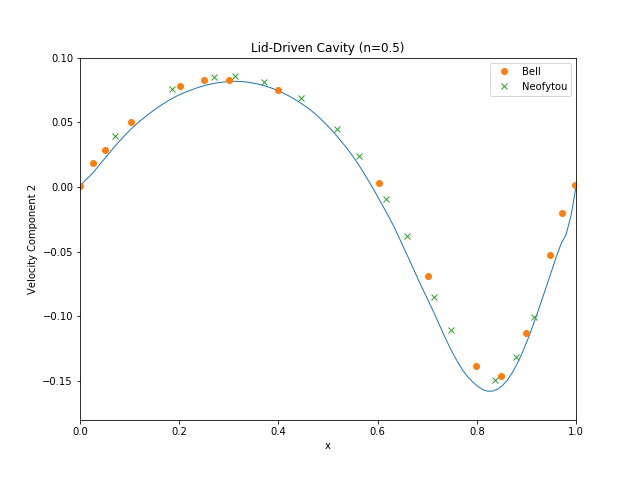}
\par\end{centering}
\caption{\label{fig:lid-driven-cavity-n=00003D0.5}Velocity profiles for the
Lid-Driven Cavity Test under our new formulation (solid line), for
a pseudoplastic with n=0.5. Slices are taken through the center of
the domain, in both axes, and compared with those of \citet{bell_p-version_1994}
and \citet{neofytou_3rd_2005}.}
\end{figure*}

\begin{figure*}
\begin{centering}
\includegraphics[width=0.5\textwidth]{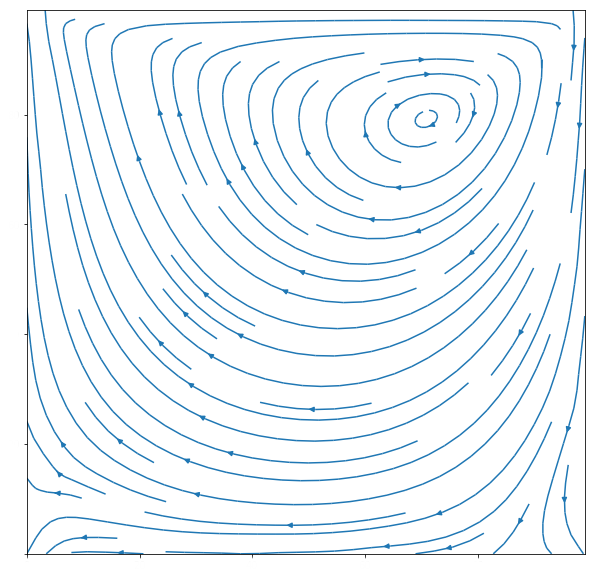}\includegraphics[width=0.5\textwidth]{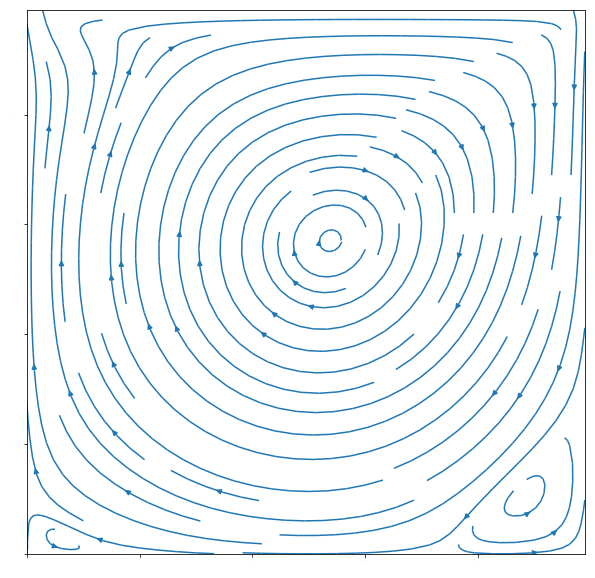}
\par\end{centering}
\caption{\label{fig:lid-driven-cavity-2d}Streamplots for the Lid-Driven Cavity
Test, for a pseudoplastic with n=0.5 (left) and a dilatant with n=1.5
(right) }
\end{figure*}

\prettyref{fig:lid-driven-cavity-n=00003D1.5} and \prettyref{fig:lid-driven-cavity-n=00003D0.5}
show the results of running the system to steady state, for $n=1.5$
and $n=0.5$, respectively. The results are compared with those of
\citet{bell_p-version_1994} and \citet{neofytou_3rd_2005}. As can
be seen, there is very good agreement for the case $n=1.5$, with
the split solver performing slightly less well for the case $n=0.5$.
The 2D streamline plots found in \prettyref{fig:lid-driven-cavity-2d}
take the characteristic forms found in the aforementioned literature.

\subsection{Elastoplastic Piston}

We now demonstrate the ability of our new numerical scheme to deal
with problems involving elastoplastic materials. This test is taken
from \citet{peshkov_theoretical_2018}, with exact solutions found
in \citet{maire_nominally_2013}.

In this test, a piston with speed $20ms^{-1}$ is driven into copper
initially at rest. An elastic shock wave develops, followed by a plastic
shock wave. The following parameters were used: $\rho_{0}=8930,c_{s}=2244,\sigma_{0}=9\times10^{7},\tau_{0}=1$.
The shock Mie-Gruneisen EOS is used for the internal energy, with
$p_{0}=0,c_{0}=3940,\Gamma_{0}=2,s=1.48$. 400 grid cells were used,
with a third order WENO method, and a CFL number of 0.7.

\begin{figure}[p]
\begin{centering}
\includegraphics[width=0.5\textwidth]{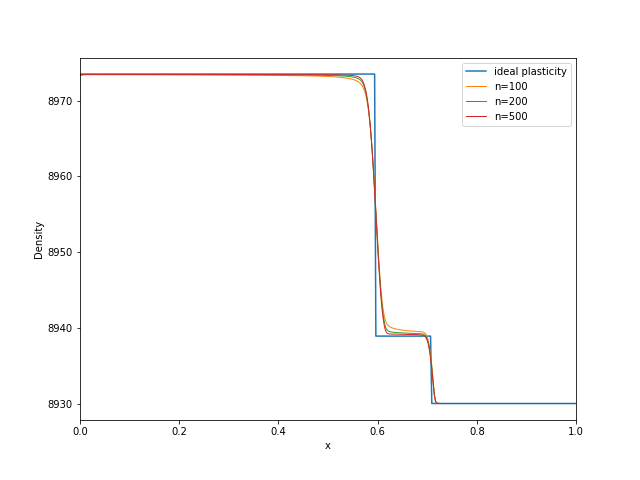}\includegraphics[width=0.5\textwidth]{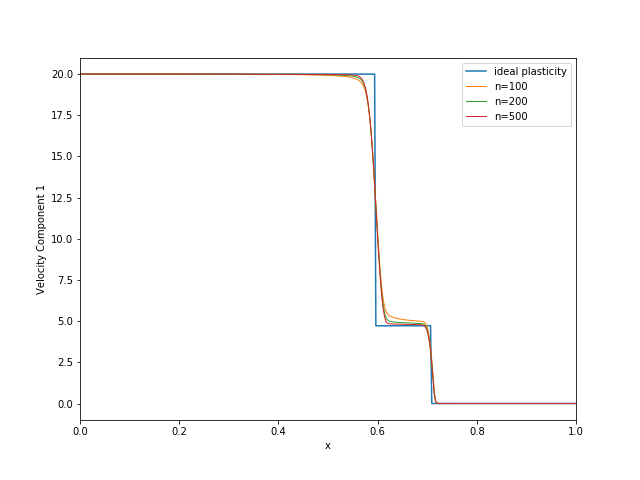}
\par\end{centering}
\caption{\label{fig:piston}Density and velocity in the elastoplastic piston
test, for various values of power-law parameter $n$}
\end{figure}

\begin{figure}[p]
\begin{centering}
\includegraphics[width=0.5\textwidth]{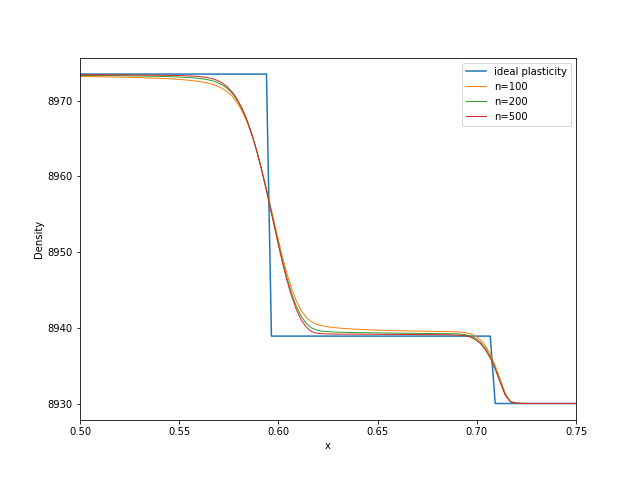}\includegraphics[width=0.5\textwidth]{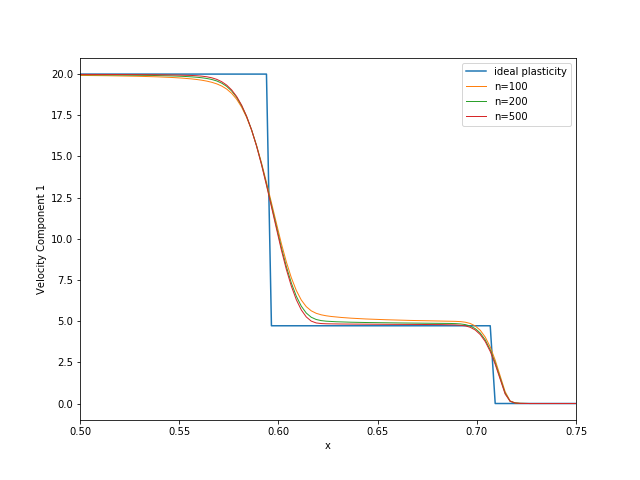}
\par\end{centering}
\caption{\label{fig:piston-1}Zoom view of density and velocity in the elastoplastic
piston test, for various values of power-law parameter $n$}
\end{figure}

\prettyref{fig:piston} and \prettyref{fig:piston-1} demonstrate
the results using the split solver for various values of $n$. These
results are compared with the exact solution to the problem under
ideal plasticity (to which the former results should converge as $n\rightarrow\infty$).
The split solver is able to cope with larger values of $n$ than those
that have been presented in \citet{peshkov_theoretical_2018}. The
results here are correspondingly closer to the ideal plasticity solution
that they approximate, than those found in the aforementioned paper.

\subsection{Cylindrical Shock}

The purpose of this test is to demonstrate the efficacy of the split
solver in multidimensional elastoplastic problems.

This test is taken from \citet{barton_eulerian_2011}. It consists
of a slab of copper, occupying the domain $\left[0,20\right]^{2}$,
initially at rest. The region $r\leq2$ is at ambient conditions,
with zero pressure. The region $r>2$ is at raised pressure $10^{10}$
and temperature $600$.

\begin{figure}[p]
\begin{centering}
\includegraphics[width=0.5\textwidth]{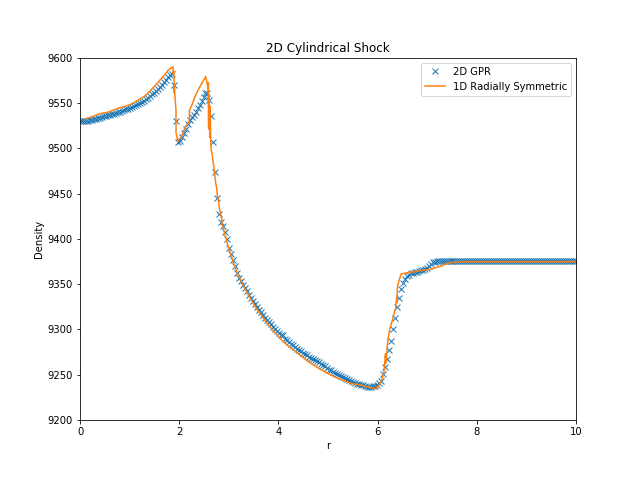}\includegraphics[width=0.5\textwidth]{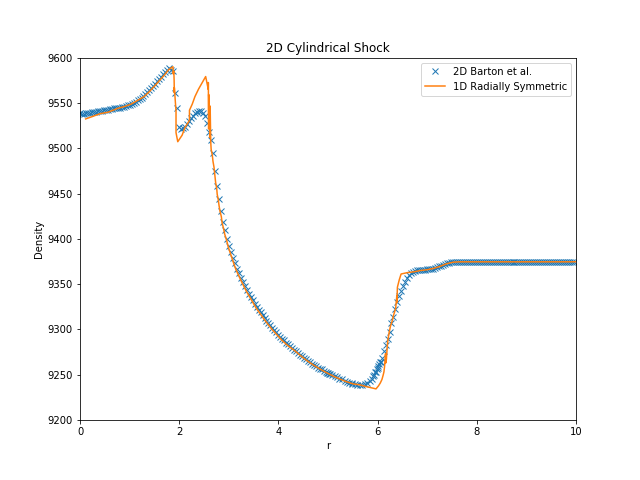}
\par\end{centering}
\caption{\label{fig:cylindrical-shock-=0003C1}1D density profiles for the
2D Cylindrical Shock Test, comparing the GPR model with split solver
(left) to the results from \citet{barton_eulerian_2011} (right)}
\end{figure}

\begin{figure}[p]
\begin{centering}
\includegraphics[width=0.5\textwidth]{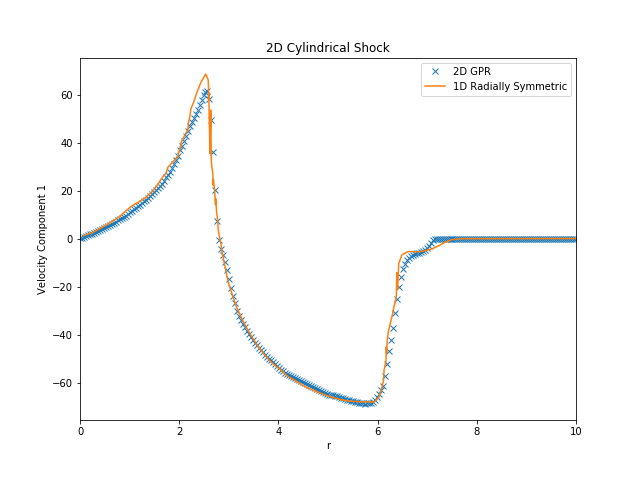}\includegraphics[width=0.5\textwidth]{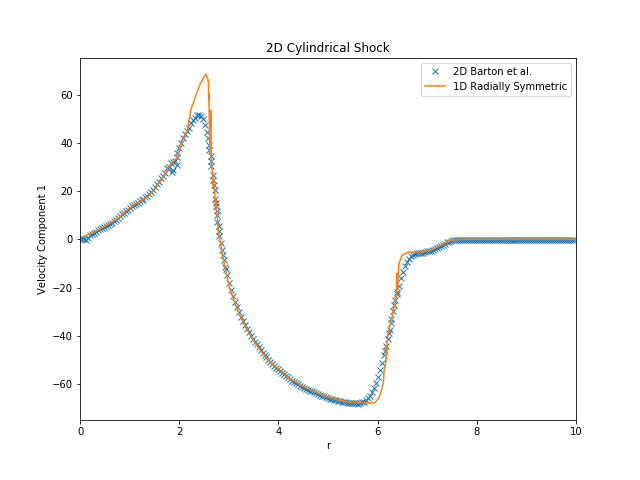}
\par\end{centering}
\caption{\label{fig:cylindrical-shock-v}1D velocity profiles for the 2D Cylindrical
Shock Test, comparing the GPR model with split solver (left) to the
results from \citet{barton_eulerian_2011} (right)}
\end{figure}

\begin{figure}[p]
\begin{centering}
\includegraphics[width=0.5\textwidth]{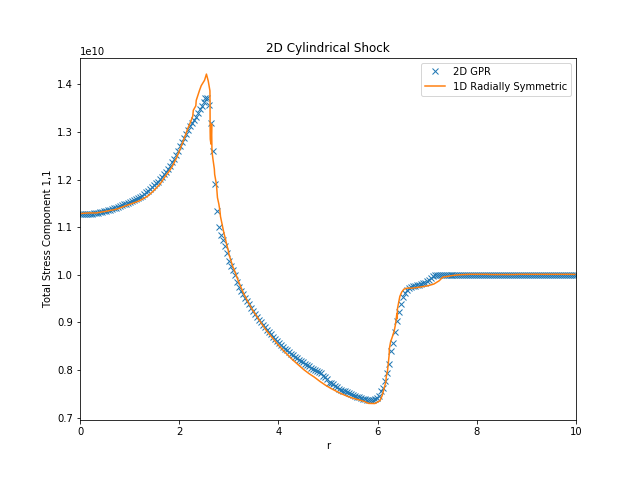}\includegraphics[width=0.5\textwidth]{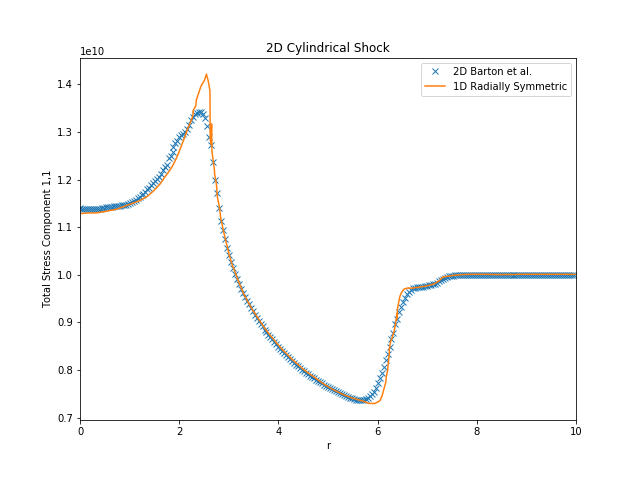}
\par\end{centering}
\caption{\label{fig:cylindrical-shock-=0003A3}1D stress tensor profiles for
the 2D Cylindrical Shock Test, comparing the GPR model with split
solver (left) to the results from \citet{barton_eulerian_2011} (right)}
\end{figure}

\begin{figure}[p]
\begin{centering}
\includegraphics[width=0.5\textwidth]{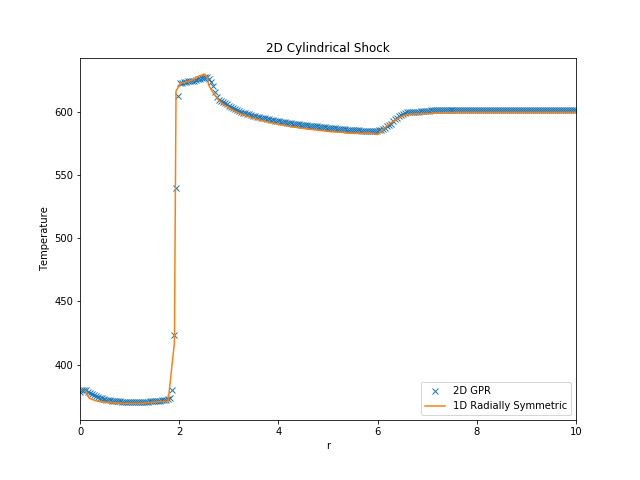}\includegraphics[width=0.5\textwidth]{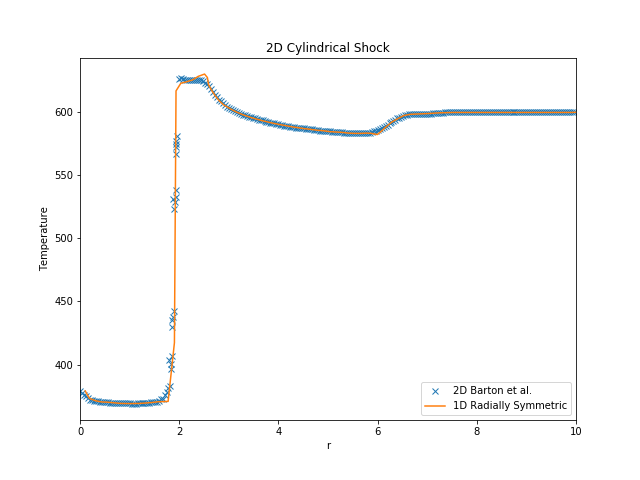}
\par\end{centering}
\caption{\label{fig:cylindrical-shock-T}1D temperature profiles for the 2D
Cylindrical Shock Test, comparing the GPR model with split solver
(left) to the results from \citet{barton_eulerian_2011} (right)}
\end{figure}

\begin{figure}[p]
\begin{centering}
\includegraphics[width=0.5\textwidth]{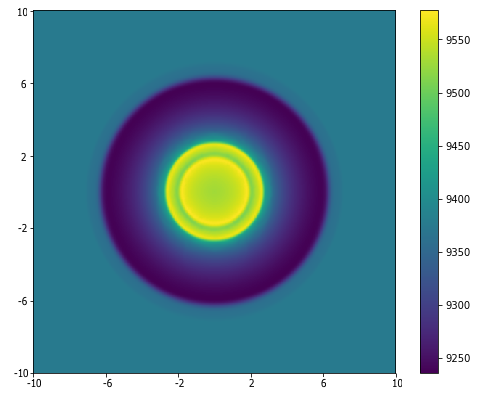}\includegraphics[width=0.5\textwidth]{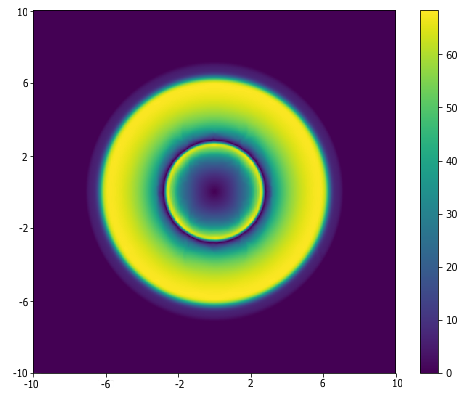}
\par\end{centering}
\caption{\label{fig:cylindrical-shock-2}2D plots of density and speed for
the Cylindrical Shock Test}
\end{figure}

The simulation is run to time $t=10^{-5}$, on a grid of shape $500\times500$.
A fourth order WENO scheme is used, with a CFL number of 0.8. The
resulting radial density, velocity, stress tensor, and temperature
profiles are given in \prettyref{fig:cylindrical-shock-=0003C1},
\prettyref{fig:cylindrical-shock-v}, \prettyref{fig:cylindrical-shock-=0003A3},
\prettyref{fig:cylindrical-shock-T}, and 2D heatmaps for density
and speed are given in \prettyref{fig:cylindrical-shock-2}.

The results are compared with those of the 1D radially-symmetric scheme
found in \citet{barton_eulerian_2011}, which are in turn compared
with the 2D results from the same publication. As can be seen, the
2D results computed using the new split solver for the GPR model more
closely match the 1D radially-symmetric results than the 2D results
from the aforementioned publication, with the spikes in both variables
around $r=2$ and the wave around $r=6$ being more accurately resolved.
Additionally, the temperature jump around $r=2$ is more sharply resolved.

\section{Conclusions\label{sec:Conclusions}}

In summary, a formulation for modeling power-law dilatants and pseudoplastics
under the GPR model has been presented. A new numerical method - based
on an operator splitting, combined with some analytical results -
has also been presented for solving this version of the GPR model,
and this numerical method has been applied also to the case of elastoplastic
solids under a power-law plasticity model. It has been demonstrated
through numerical simulation that the modified GPR formulation is
able to accurately describe the evolution of non-Newtonian fluids,
and the new numerical scheme has been shown to be an effective method
by which to solve this system, and the existing corresponding system
for elastoplastic solids.

Under circumstances in which the flow is compressed heavily in one
direction relative to the other directions, it should be noted that
the linearization assumption \eqref{eq:Assumption} used to derive
the approximate analytical solver may break down. As discussed in
\citet{jackson_fast_2017}, this is due to the fact that one of the
singular values of the distortion tensor will be much larger than
the others, and the mean of the squares of the singular values will
be distant to the geometric mean. The subsequent linearization of
the ODE governing the mean of the singular values will then fail.
It should be noted that none of the situations covered in this study
presented problems for the approximate analytical solver, and situations
which may be problematic are in some sense unusual. In any case, a
stiff ODE solver can be used to solve the systems \eqref{eq:DistortionODE},
\eqref{eq:ThermalODE} if necessary, and so this method is still very
much usable in these situations, albeit slightly slower.

As detailed in \citet{leveque_study_1990}, solvers based on a temporal
splitting suffer from a lack of spatial resolution in evaluating the
source terms. Thus, it should be noted that the operator splitting
method presented here may suffer from the incorrect speed of propagation
of discontinuities on regular, structured grids. This issue can be
rectified, however, by the use of some form of shock tracking or mesh
refinement, as noted in the cited paper. \citet{dumbser_finite_2008}
note that operator splitting-based methods can result in schemes that
are neither well-balanced, nor asymptotically consistent. The extent
to which these two conditions are violated by this method \textendash{}
and the severity in practice of any potential violation \textendash{}
is a topic of further research. 

It should be noted that the new numerical scheme presented in this
study is trivially parallelizable on a cell-wise basis. Thus, given
a large number of computational cores, deficiencies in this method
in terms of its order of accuracy may be overcome by utilizing a larger
number of computational cells and cores. The number of grid cells
that can be used scales roughly linearly with number of cores, at
constant time per iteration.

\section{References}

\bibliographystyle{elsarticle-harv}
\bibliography{28_Users_hari_Git_private_phd_papers_Working_A____and_Plastic_Solids_under_the_GPR_Model_refs}

\section{Acknowledgments}

The authors acknowledge financial support from the EPSRC Centre for
Doctoral Training in Computational Methods for Materials Science under
grant EP/L015552/1.
\end{document}